\ifdefined\pdfminorversion\pdfminorversion=7\fi
\documentclass[11pt,a4paper]{article}
\usepackage[T1]{fontenc}
\usepackage[utf8]{inputenc}
\usepackage[english]{babel}
\usepackage{lmodern}
\usepackage[a4paper,margin=24mm]{geometry}
\usepackage{microtype}
\usepackage{amsmath,amssymb}
\usepackage{graphicx}
\usepackage{booktabs,tabularx,array,multirow}
\usepackage{enumitem}
\usepackage{caption}
\usepackage{placeins}
\usepackage[numbers,sort&compress]{natbib}
\usepackage{tikz}
\usetikzlibrary{arrows.meta,positioning}
\usepackage[hidelinks]{hyperref}
\hypersetup{pdftitle={Multimodal Concert Video Editing: System Design and Exploratory Evaluation},pdfauthor={Lennart Schiweck et al.}}
\setlist{nosep,leftmargin=*}

\newcommand{\clip}{\operatorname{clip}}
\newcommand{\NA}{\textemdash}
\newcolumntype{Y}{>{\raggedright\arraybackslash}X}

\title{Multimodal Concert Video Editing: System Design and Exploratory Evaluation}
\author{Lennart Schiweck, Adrian Kempf, Michelle Retig,\\Moritz Feucht, Tim Mangliers and Benjamin Himpel\\[5pt]
\small Department of Informatics, Reutlingen University\\
\small 72762 Reutlingen, Germany\\[3pt]
\small Corresponding author: Tim Mangliers\\
\small \href{mailto:Tim.Mangliers@Reutlingen-University.de}{Tim.Mangliers@Reutlingen-University.de}}
\date{}

\begin{document}
\maketitle
\begin{abstract}
Editing multi-camera concerts requires coordinated decisions about musical relevance, visual quality and cut timing. We present a modular framework that combines instrument activity from multitrack audio, visible instruments and image sharpness with beat-based timing and probabilistic camera selection. An interpretable scoring model links these components to the choice of views over complete shot intervals. Component tests and an exploratory expert study on a four-camera concert yield mixed assessments across the editing variants. The findings motivate extending local audiovisual relevance with contextual and sequence-level criteria. The architecture provides a basis for testing these extensions and studying how component behaviour affects the quality of complete edits.
\end{abstract}

\section{Introduction}
Multi-camera concert editing requires deciding when to cut and which view to show. Editors coordinate musical activity, image quality, framing and the progression of a sequence. A clear view of an active instrument can be useful, but so can a reaction or an interaction elsewhere in the ensemble. Automating this workflow therefore requires more than ranking isolated images.

Automatic camera management has been studied in lecture environments \citep{Liu2001}, concert mashups \citep{Shrestha2010} and social-camera footage \citep{Arev2014}. Concert-directing systems combine auditory and visual instrument information \citep{Wang2023}; learned approaches address next-view selection and cut prediction \citep{Jimenez2022,GonzalbezBiosca2025}. Temporal and candidate-view context are also modelled in multi-camera editing \citep{Rao2022,CabacasMaso2026}, while theatre systems combine camera selection and shot-size control \citep{Stoll2023}.

We present a modular framework that separates audio analysis, visual analysis, camera scoring and temporal selection. Named audio stems provide instrument activity and roles; visual detections describe instrument visibility, and image-gradient statistics characterise sharpness. An explicit utility combines these signals, aggregates them over a prospective shot interval and controls a probabilistic camera selector. Beat and tempo estimates determine candidate cut boundaries independently of view selection.

The contributions are threefold: a transparent multimodal architecture for concert editing; an explicit mapping from role and quality cues to interval-level camera choices; and a component and expert evaluation that motivates targeted architectural extensions. The study compares a human reference, random camera selection and two score-driven variants. Its mixed results inform further work on shot suitability, contextual reactions and sequence structure without treating component accuracy as a substitute for complete-edit quality.

\section{Related work}
\label{sec:related}
Wang et al.~\citep{Wang2023} provide a close comparison to the explicit-score design: their system already combines audible--visible instrument matching, blur and shot-scale information. Table~\ref{tab:related} compares the audio inputs, scoring, timing and selection mechanisms.

\begin{table}[htbp]\centering\small
\caption{Concert-directing comparison with Wang et al. (2023). The table compares system designs rather than performance on a common dataset.}
\label{tab:related}
\begin{tabularx}{\textwidth}{@{}p{.18\textwidth}YY@{}}\toprule
Axis & Wang et al. (2023) & Our system\\\midrule
Audio & Instrument classification from mixed audio. & Named stems; relative energy for primary/secondary roles.\\
Instrument matching & Audio--visual instrument agreement contributes when permitted by shot scale. & Explicit audible--visible role bonuses and missing/inaudible-label penalties.\\
Image criteria & Blur and shot-scale information. & Fixed gradient blur thresholds and box-area roles; no framing optimisation.\\
Temporal handling & Downbeat information and an estimated switching rate. & Every-fourth-beat grid and tempo-based target duration; conservative full-interval scoring.\\
Cut positions & Downbeat-based timing. & Next grid point after a tempo-dependent target duration.\\
Camera selection & Highest eligible score under shot-scale transition rules. & Proportional sampling with no immediate repeat and uniform fallback.\\
Evaluation & Automatic/manual clips rated by 16 participants. & Four full edits, three raters, random and annotation-driven diagnostics, original comments.\\
\bottomrule\end{tabularx}\end{table}

Shrestha et al.~\citep{Shrestha2010} combine perceptual image quality, content diversity and cut-point suitability in concert mashups. Their 40-participant evaluation already compares automatic, manual and random edits, so our baseline follows an established strategy. Daniyal and Cavallaro~\citep{Daniyal2011} model finite-horizon camera scheduling with content quality and switching costs; our no-repeat sampler does not optimise a sequence-level objective. Stoll et al.~\citep{Stoll2023} combine rule-based camera selection and cropping for four-perspective theatre recording, including audiovisual speaker assignment. Their comparison of four-perspective and single-perspective edited versions holds cut times, shot lengths and shot sizes fixed. Our system selects existing views without generating tighter framing.

Learned concert-editing approaches are also direct comparators. Jim{\'e}nez et al.~\citep{Jimenez2022} learn visual next-shot selection from professionally edited concert videos. Their human study assesses static image choices. Learned next-view selection is therefore an established approach. Gonz{\'a}lbez-Biosca et al.~\citep{GonzalbezBiosca2025} separate cut-time prediction from view selection: audio and temporal features inform the former, and learned visual representations inform the latter. Their classical-concert evaluation uses pseudo-labelled candidates derived from existing edits and reports cut-prediction and retrieval measures. These are different outcomes from perceived quality of a complete realised edit.

Rao et al.~\citep{Rao2022} introduce the TVMCE benchmark with synchronised multi-camera material including concerts and model both past frames and candidate views. Cabacas-Maso et al.~\citep{CabacasMaso2026} further distinguish history and candidate encoding through a dual-transformer architecture with cross-attention. Such approaches explicitly represent context beyond our no-repeat sampler, but their reference-decision agreement measures do not by themselves establish perceptual quality of our concert edits. 

Adjacent editorial approaches make broader objectives explicit. Galvane et al.~\citep{Galvane2015} optimise shot sequences in 3D animation using action, continuity and rhythm costs, with subjective comparisons of degraded cost variants. EditIQ \citep{Girmaji2025EditIQ} combines dialogue interpretation, saliency and speaker cues with cinematic constraints for virtual views from wide-angle recordings; its user study includes emotions and reactions. EmotionCast \citep{Zhang2024EmotionCast} additionally studies affective cues in stage-program directing. These works motivate testing editorial context and sequence-level criteria, while their different materials and designs prevent transferring their performance claims to this pilot. The concert evaluation motivates testing these ideas for instrumental performance and ensemble interaction.

The emphasis of our framework is an explicit, modular connection between instrument roles, quality cues, shot timing and camera choice. Component measurements and complete-edit assessment provide complementary views of this design.

\section{System overview and material}
\label{sec:material}
\begin{figure}[htbp]\centering
% Editable TikZ fragment for \input inside a figure environment.
% Requires tikz with arrows.meta. Native size is below 162 x 90 mm.
% All labels are 8.5 pt; use at native size.
\begin{tikzpicture}[
  x=1mm,y=1mm,
  font=\fontsize{8.5}{10}\selectfont,
  sys block/.style={draw=black!65,fill=black!2,rounded corners=1mm,
    line width=.45pt,align=center,inner xsep=2mm,inner ysep=1.3mm},
  sys input/.style={sys block,fill=white,minimum width=28mm,
    text width=24mm,minimum height=14mm},
  sys analysis/.style={sys block,minimum width=40mm,text width=36mm,
    minimum height=14mm},
  sys utility/.style={sys analysis,fill=blue!5!white},
  sys decision/.style={sys block,fill=blue!5!white,
    minimum width=26mm,text width=22mm},
  sys flow/.style={-{Latex[length=1.7mm,width=1.1mm]},
    draw=black!70,line width=.55pt},
  sys alternative/.style={sys flow,dashed,draw=black!60}
]
% The timing waveform is the mono mix of the named audio stems.
\draw[draw=black!55,fill=black!1,rounded corners=1.5mm,line width=.5pt]
  (0.5,14.5) rectangle (31.5,65);
\node[font=\fontsize{8.5}{10}\selectfont\bfseries,inner sep=0]
  at (16,60) {Audio inputs};
\node[sys input] (sys-stems) at (16,47)
  {\textbf{Named audio}\\\textbf{stems}};
\node[sys input] (sys-wave) at (16,25)
  {\textbf{Mono timing}\\\textbf{waveform}};

\node[sys input,minimum height=17mm] (sys-cameras) at (16,78)
  {\textbf{Four cameras}\\(synchronised)};
\node[sys analysis,minimum height=17mm] (sys-visual) at (59,78)
  {\textbf{Visible roles + blur}\\Instrument boxes / area;\\image-gradient thresholds};
\node[sys analysis] (sys-roles) at (59,47)
  {\textbf{Activity / audible roles}\\Smoothed relative RMS};
\node[sys analysis] (sys-tempo) at (59,25)
  {\textbf{Beat / tempo analysis}\\Every-fourth-beat grid};

\node[sys utility,minimum height=22mm] (sys-score) at (108,65)
  {\textbf{Per-second utilities}\\Audible--visible matches,\\penalties and quality gate};
\node[sys analysis] (sys-boundaries) at (108,25)
  {\textbf{Shot boundaries}\\Tempo-based target duration};
\node[sys analysis,dashed,fill=white,minimum height=13mm]
  (sys-labels) at (108,42)
  {\textbf{Reference labels}\\Alternative input;\\optional pan/clapping};

\node[sys decision,minimum height=20mm] (sys-interval) at (148,65)
  {\textbf{Interval utility}\\Lowest 10\% mean;\\clipped at zero};
\node[sys decision,minimum height=20mm] (sys-select) at (148,25)
  {\textbf{Camera selection}\\No repeat;\\draw $p_c\propto q_c$;\\uniform fallback};
\node[sys decision,minimum height=11mm,fill=black!7] (sys-edit) at (148,5)
  {\textbf{Assembled edit}\\Time / camera};

\draw[sys flow] (sys-cameras) -- (sys-visual);
\draw[sys flow] (sys-stems) -- (sys-roles);
\draw[sys flow] (sys-stems.south) -- node[right,inner sep=.8mm] {mix} (sys-wave.north);
\draw[sys flow] (sys-wave) -- (sys-tempo);
\draw[sys flow] (sys-visual.east) -- ([yshift=4mm]sys-score.west);
\draw[sys flow] (sys-roles.east) -- ([yshift=-5mm]sys-score.west);
\draw[sys alternative] (sys-labels.north) -- (sys-score.south);
\draw[sys flow] (sys-score) -- (sys-interval);
\draw[sys flow] (sys-tempo) -- (sys-boundaries);
\draw[sys flow] (sys-boundaries.east) -- (132,25)
  -- (132,60) -- ([yshift=-5mm]sys-interval.west);
\draw[sys flow] (sys-interval) -- (sys-select);
\draw[sys flow] (sys-select) -- (sys-edit);

\draw[draw=black!70,line width=.55pt] (1,6) -- (8,6);
\node[anchor=west,align=left,inner sep=0] at (10,6)
  {Analysis and selection paths};
\draw[draw=black!60,line width=.55pt,dashed] (1,1) -- (8,1);
\node[anchor=west,align=left,inner sep=0] at (10,1)
  {Reference-label alternative};
\end{tikzpicture}
\caption{Modular concert-editing architecture. Named stems supply activity analysis and are mixed into a mono waveform for timing. Visual roles and sharpness combine with audible roles in a camera utility; shot boundaries define the aggregation intervals. Camera selection produces the edit. Reference annotations form an alternative label input for the annotation-driven evaluation variant.}
\label{fig:architecture-overview}\end{figure}
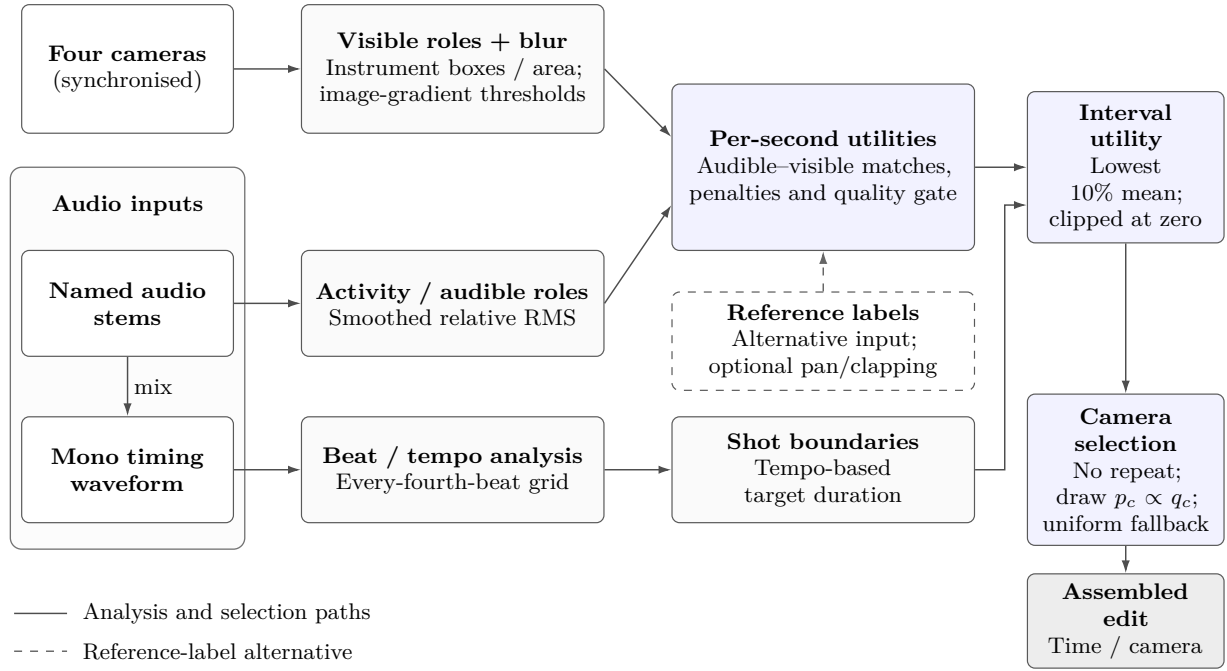

The framework takes four synchronised camera streams and concert audio as input. Audio processing has two complementary branches: named instrument stems supply activity and audible roles, while their mono mix supplies beat and tempo estimates. The visual branch uses a pretrained D-FINE object detector with retained instrument categories from Objects365 \citep{Peng2024,objects365} and image-gradient statistics for sharpness. These signals feed the camera utility, while the timing branch defines intervals over which views are evaluated. Figure~\ref{fig:architecture-overview} gives the complete processing flow.

The evaluation uses a recording of the Uli Gutscher Quartet performing at JAZZclub IN DER MITTE e.V. in Reutlingen on 11 January 2025. The camera arrangement comprises an audience-oriented view, two close-up views and a wide view (Table~\ref{tab:cameras}). Component measurements cover the first 480\,s; the expert study uses four complete edits of approximately 7\,min\,48\,s. The annotated audio vocabulary comprises Bass, Drums, Guitar, Piano, Speech and Trombone. Labels describe audible and visible roles, blur and, for the annotation-driven variant, additional pan and clapping cues.

\begin{table}[htbp]
\centering\small
\caption{Camera mapping and blur reference supports. $n$ is the number of one-second evaluation samples used in the blur evaluation. Supports are ordered as not blurred / partially blurred / fully blurred.}
\label{tab:cameras}
\begin{tabularx}{\textwidth}{@{}lYrr@{}}
\toprule
Camera & View & $n$ & Class supports\\
\midrule
CAM1 & Audience-oriented view & 481 & 7 / 474 / 0\\
CAM2 & Close-up view & 480 & 54 / 386 / 40\\
CAM3 & Close-up view & 481 & 4 / 453 / 24\\
CAM4 & Wide view & 481 & 481 / 0 / 0\\
\bottomrule
\end{tabularx}
\end{table}

The prediction-driven variant combines estimated audio roles, visible roles and blur. The annotation-driven variant replaces these inputs with reference labels and also includes pan/clapping cues. It therefore tests annotation-driven camera scoring, rather than a detector-only substitution. Detailed algorithms and parameters are given in Appendix~\ref{app:implementation}.

\section{Camera scoring and selection}
\label{sec:methods}
The system translates audible and visible instrument roles into per-second camera utilities, combines them with image quality and aggregates them over each candidate shot. The selection module samples one camera per interval while preventing immediate camera repetition.

\subsection{Local utility}
\label{sec:utility}
At integer second $t$, audible primary/secondary labels are matched against visible primary/secondary labels in each camera. Let $R_c(t)$ sum these matches, with bonuses 4, 1, 1 and 0.5 for primary--primary, primary--secondary, secondary--primary and secondary--secondary matches. The penalty is
\begin{equation}
P_c(t)=0.9|A_p(t)\setminus V_c(t)|+0.5|A_s(t)\setminus V_c(t)|
+0.75|V_c(t)\setminus A(t)|,
\end{equation}
where $A_p,A_s$ are audible-role sets and $A,V_c$ are the audible and camera-visible unions. The score is
\begin{equation}
S_c(t)=\operatorname{round}_2\bigl([4+R_c(t)-P_c(t)]Q_c(t)\bigr).
\label{eq:score}
\end{equation}
The quality multiplier $Q_c$ is zero for a fully blurred or recognised pan label and otherwise one; partial blur is not penalised. Pan cues are used by the annotation-driven variant.

\subsection{Timing, interval utility and selection}
\label{sec:timing}\label{sec:selection}
The timing module places boundaries on every fourth tracked beat and samples target durations of 10--15, 6--10 or 3--6\,s for slow, middle or fast tempo classes. This assumes four beats per bar and does not establish true downbeats or musical phrasing. 

For a prospective interval, scores are aggregated conservatively. If $S_{c,(1)}\leq\cdots\leq S_{c,(n)}$ are its integer-bin scores,
\begin{equation}
q_c=\max\left(0,\frac{1}{m}\sum_{j=1}^{m}S_{c,(j)}\right),\qquad
m=\max(1,\lceil0.10n\rceil).
\label{eq:tail}
\end{equation}
For at most ten bins this is the minimum clipped at zero. Given previous camera $i$, proportional sampling selects $j\ne i$ with
\begin{equation}
p_j=\frac{q_j}{\sum_{k\ne i}q_k},\qquad p_i=0,
\label{eq:selection}
\end{equation}
when the denominator is positive; otherwise the fallback is uniform over the other three cameras. A zero score is therefore not a hard veto. This is a no-repeat constraint rather than a sequence model of editorial grammar. Full-interval scoring and centred audio analysis use future context; the present implementation operates offline.

\section{Component performance}
\label{sec:components}
Audio-role estimation reaches support-weighted F1 scores of 0.654 for primary and 0.601 for secondary roles. Performance varies across instruments, and blur classification is strongly camera-dependent. These differences affect how the corresponding cues can guide camera choice. Appendix~\ref{app:component-details} reports the per-class audio and blur results, sampling definitions, role timelines and camera-score plots.

\section{Expert evaluation}
\label{sec:experts}
\subsection{Study design}
Four edits of the same recording are compared: V1 is a student-produced human reference under expert supervision; V2 uses random camera selection with beat-based timing; V3 uses predicted camera scores; and V4 uses annotation-driven scores. Three experts from different media fields assessed the complete versions without being told the editing method.

The questionnaire requests an overall publication/broadcast-readiness grade and detail ratings for cut timing, transitions, picture quality/sharpness and coherence/aesthetic impression, together with free-text comments. Overall grades use 1 = very good and 5 = poor. Conflicting direction anchors make the detail scales ambiguous, so only the overall grades are interpreted directionally; the detail responses are reported without reversal in Appendix~\ref{app:rawratings}. Appendix~\ref{app:questionnaire} describes the questionnaire dimensions.

\subsection{Cut timing and shot durations}
\label{sec:observedcuts}
The four edits use the same audio. V1 contains 35 internal camera changes; V2--V4 each contain 65. Corresponding automatic cut boundaries differ by at most one frame: V3 and V4 coincide, while V2 is one frame earlier. Thus the automatic variants have effectively the same pacing, allowing their view choices to be compared at similar cut timing.

The clip comprises an opening announcement, music and final applause. We use approximate windows of 0--80\,s, 80--460\,s and 460--467.7\,s. Table~\ref{tab:shots} reports the duration of displayed segments over the full clip and the music window. Shots crossing a window boundary are clipped for the summary, without counting an additional camera change. Appendix~\ref{app:cuts} gives the definitions and additional phase summaries.

\begin{table}[htbp]
\centering\small
\caption{Shot-duration summaries from the complete rated videos and the approximately 380-s music window (80--460\,s). Durations are in seconds. Cuts are internal camera changes in the window; $n$ is the number of positive displayed segments, including clipped boundary segments. SD is the sample SD, not uncertainty across concerts. V2--V4 are listed separately, not pooled.}
\label{tab:shots}
\begin{tabular}{@{}llrrrrr@{}}
\toprule
Variant & Window & Cuts & $n$ & Mean & Median & SD\\
\midrule
V1 & Full & 35 & 36 & 12.99 & 6.94 & 19.70\\
V2 & Full & 65 & 66 & 7.09 & 6.72 & 2.27\\
V3 & Full & 65 & 66 & 7.09 & 6.72 & 2.27\\
V4 & Full & 65 & 66 & 7.09 & 6.72 & 2.27\\
\midrule
V1 & Music & 34 & 35 & 10.86 & 7.14 & 9.68\\
V2 & Music & 55 & 56 & 6.79 & 6.67 & 2.18\\
V3 & Music & 55 & 56 & 6.79 & 6.67 & 2.18\\
V4 & Music & 55 & 56 & 6.79 & 6.67 & 2.18\\
\bottomrule
\end{tabular}
\end{table}

V1 begins with an uninterrupted 116.72-s shot, giving longer and more variable durations than the automatic edits. In the music window its mean is 10.86\,s, compared with 6.79\,s for V2--V4. The automatic versions also make nine cuts during the announcement, whereas V1 makes none. These statistics describe pacing; musical and editorial fit are assessed separately in the expert feedback.

\subsection{Overall grades}
\begin{table}[htbp]\centering\small
\caption{Complete overall grades, 1 = very good and 5 = poor. Mean, median and sample SD summarise three ordinal responses descriptively; they do not estimate uncertainty across concerts.}
\label{tab:ratings}
\begin{tabular}{@{}lrrrrrr@{}}\toprule
Variant & E1 & E2 & E3 & Mean & Median & SD\\\midrule
V1: student reference &3&3&3&3.00&3&0.00\\
V2: random &4&3&1&2.67&3&1.53\\
V3: predicted scores &5&4&3&4.00&4&1.00\\
V4: annotated scores &4&4&1&3.00&4&1.73\\
\bottomrule\end{tabular}\end{table}

The assessments are mixed across experts and variants. V1 receives consistent overall grades, while V2 and V4 show greater disagreement. All three experts grade V3 less favourably than V2. V4 improves on V3 for two experts and ties it for the third, but does not receive a better grade than V2. In this recording, score-driven selection therefore provides no overall advantage over the random baseline. The individual grades remain important: equal grades do not necessarily express equal preference, as E3 explicitly prefers V2 despite giving V2 and V4 the same grade.

\subsection{Written feedback}
The free-text responses are grouped descriptively by editorial theme. Table~\ref{tab:themes} retains positive, critical and differing assessments rather than reducing them to a single consensus. The analysis concerns the experts' expressed judgements; it does not assign an internal technical cause to each reported shot. Appendix~\ref{app:qualitative} describes the grouping procedure.

\begin{table}[htbp]\centering\small
\caption{Themes in the written expert feedback. Locators identify expert, video and comment field.}
\label{tab:themes}
\begin{tabularx}{\textwidth}{@{}>{\raggedright\arraybackslash}p{.20\textwidth}Y>{\raggedright\arraybackslash}p{.24\textwidth}@{}}\toprule
Concern & Observation and counterpoint & Example locators\\\midrule
Musical action and view choice & E1 describes V3's cut positions as matching V2 but its chosen views as worse, often missing the musician carrying the theme. & E1/V3: cut comment, remarks\\
Appropriate moments & Nose blowing and people standing or leaving attract criticism in several variants; this is not exclusive to V3. & E1/V2: cut comment; E1/V4: cut comment; E3/V3: impression comment\\
Framing and view diversity & E2 repeatedly requests better framing and close-ups; E3 criticises the detail-poor wide view. These comments do not prove that better raw alternatives were unavailable. & E2/V1--V4: remarks; E3/V1: sharpness comment\\
Interaction and reactions & E3 praises reactions, laughter and musical responses in V2. Such views can be valuable without showing only the leading instrument. & E3/V2: impression comment, remarks\\
Different pacing criteria & E1/E2 find V2's pattern schematic or unmotivated, while E3 finds it coherent and dynamic. & E1/V2: cut, transition comments; E2/V2: cut comment; E3/V2: cut comment\\
Contextual audience value & E1/E2 criticise V3's audience choices; E3 also values some audience reactions while rejecting unmotivated views. & E1/V3: remarks; E2/V3: cut comment; E3/V3: cut and impression comments\\
\bottomrule\end{tabularx}\end{table}

The feedback highlights a distinction between technically usable images and editorially useful moments. It also values reactions and band interaction alongside the currently prominent instrument. Criticism of incidental audience views coexists with praise for motivated reactions, so the comments do not support a general audience-view prohibition.

The complete edits and an illustrative side-by-side excerpt accompany the study.\footnote{Full variants: \url{https://drive.google.com/drive/folders/1qKtlRt0Fae0aA4U3sTLfgcf3DPBDQV6c}; excerpt: \url{https://drive.google.com/file/d/1vLJxgV4vO4KYswahPImz7GACkTWG7hM4/view}.}

\section{Discussion}
\subsection{Relating system components to editorial assessment}
The evaluation relates component-label agreement, temporal characteristics and expert assessment of the complete sequence. The automatic variants share effectively the same measured boundaries, while their overall grades and reported view choices differ. This narrows the interpretation towards camera selection, cue content and sequence behaviour; it does not identify a single causal component. V4's improvement on V3 for two raters also cannot isolate recognition accuracy, because annotation source, available cues and potentially random draws change together.

The feedback specifies what a useful view means to these raters: coverage of musical action, acceptable framing and moments, and motivated reactions or ensemble interaction. These criteria span different architectural interfaces. Table~\ref{tab:diagnosis} links the reported concerns to operations that can influence selection and to changes that can be tested. The feedback motivates each link; it does not show that the named operation caused a particular shot.

\subsection{Tracing component outputs into camera choices}
\begin{table}[htbp]\centering\small
\caption{Component-to-evaluation links and proposed architectural tests. Operational consequences follow from the scoring and selection rules; feedback locators identify reported outcomes. Proposed improvements have not been evaluated.}
\label{tab:diagnosis}
\begin{tabularx}{\textwidth}{@{}>{\raggedright\arraybackslash}p{.29\textwidth}>{\raggedright\arraybackslash}p{.28\textwidth}Y@{}}\toprule
Component and score pathway & Expert assessment & Architectural change and isolated test\\\midrule
Audio roles: smoothed relative energy changes match bonuses and missing-label penalties. & E1/V3 remarks report missing musical action; primary-role F1 is 0.654. & Activity and musical relevance could be separated; smoothing or relevance input could vary independently, with score/choice changes logged.\\
Visible roles: box area supplies prominence, without framing, action suitability or reaction context. & E2/V1--V4 remarks criticise framing; E1/V4 cut comment reports an unsuitable moment; E3/V2 remarks praise reactions. & Shot eligibility could be distinguished from musical/contextual relevance, with each tested using independently annotated inputs before introducing a new detector.\\
Quality map: partial blur and not blurred both have multiplier one. & CAM1 has 474 partial-blur reference samples but is predicted not blurred; E1/V2 and E3/V3 sharpness comments mention focus. & Correcting those labels alone has no score effect. A calibrated quality map could be tested separately from label accuracy.\\
Interval/policy: short intervals use their minimum score clipped at zero; zero utility is not a veto under fallback. & E1/V3 cut comment reports poorer views at V2-like cut times; the internal reason is unobserved. & Eligibility, aggregation and sampling could be varied one at a time, with complete decision logs.\\
Timing: a beat/tempo heuristic also covers speech; automatic pacing is matched. & Nine announcement cuts; E2/V2 and V4 cut comments criticise speech timing; E3/V2 cut comment praises rhythm. & A speech/music/applause state could support timing tests separately from view selection.\\
Cue interface: V4 adds pan/clapping labels alongside reference roles. & V4 improves on V3 for two raters but does not outgrade V2. & Label sources could be compared under cue parity, with optional cues then tested as a separate factor.\\
\bottomrule\end{tabularx}\end{table}

An important example is the difference between measurement error and operational influence. CAM1's partial/not-blurred disagreement produces poor classification metrics, but both states have $Q_c=1$. Correcting only that disagreement therefore leaves its score unchanged. It cannot, by itself, explain audience-view selection. By contrast, audible-role changes enter both bonuses and penalties, and full-blur or recognised pan labels can zero a per-second score. Even then, the interval aggregate, previous camera and fallback determine whether a score change reaches a displayed view. Appendix~\ref{app:architecture-tests} gives conditional score examples and controlled comparisons for studying these effects.

\FloatBarrier
The objective also omits some distinctions the comments make. A large, sharp view of a relevant instrument need not show a suitable action, and a valuable reaction need not depict the leading instrument. Adding a larger instrument bonus would not resolve that representational issue. At the same time, neither audience presence nor a non-leading musician is inherently unsuitable: E3's positive reaction feedback argues for contextual assessment rather than a blanket penalty. Source coverage must be inspected separately, since selection cannot create an unavailable close-up.

\subsection{Proposed controlled comparisons}
A useful next version would expose separate interfaces for cue availability and reliability, candidate eligibility, musical/contextual relevance, and the sequence policy. Timing would receive an explicit speech/performance/applause state. This factorisation makes it possible to ask whether a recognised property is accurate, whether it changes the utility as intended, and whether the resulting views improve the relevant editorial judgement. It also makes disagreement informative: reaction value and pace preference can be reported per rater rather than hidden in one mean.

\textbf{Comparison 1: separating suitability and relevance.} At common decision opportunities and with unchanged remaining inputs, an initial comparison could add independently annotated eligibility information to the existing selector; a separate comparison could test contextual reaction value. Outcomes would include unsuitable-view exposure, musical-action coverage and paired sequence preference, with any changes in realised cuts caused by eligibility exceptions reported explicitly. This would first test the editorial criterion before introducing prediction error from a new detector. Action-aware and reaction-sensitive objectives have precedents in other settings \citep{Galvane2015,Girmaji2025EditIQ,Zhang2024EmotionCast}; their value for this concert remains to be measured.

\begin{samepage}
\textbf{Comparison 2: adding sequence context.} A controlled comparison would hold local scores and timing fixed, retain the same no-repeat and fallback rules, and compare selection with and without one prespecified history/transition term, such as recently shown performers or view roles. An inspectable rule would avoid initially requiring a larger learned model. Temporal-context systems \citep{Rao2022,CabacasMaso2026} and learned concert selection \citep{Jimenez2022,GonzalbezBiosca2025} offer later comparison models. Several prespecified realisations and blinded sequence judgements are needed; reference-decision agreement alone is insufficient.\par
\end{samepage}

\textbf{Comparison 3: making quality and fallback operationally explicit.} Quality could first be calibrated on development material with supported classes per camera, followed by independent tests of the quality-to-utility map and fallback policy. This comparison would distinguish a hard eligibility constraint from a soft preference. If only the current view is usable, holding it may be preferable to a mandatory switch; if no view is eligible, the interval could be flagged or an explicitly documented exception invoked. Such holds change realised cuts and dwell times, so common decision opportunities do not imply matched realised pacing. The pilot motivates these tests but establishes no improvement from them.

Phase-specific timing is a complementary extension. Speech/content boundaries should be assessed during announcements, and beat/downbeat or phrase cues during music. Its effect requires a separate timing comparison. The approximately 80-s announcement and the different rhythm judgements show why a full-clip grade and a music-window duration summary answer different questions. A consistent cue vocabulary is also necessary before changing predicted to reference labels; pan/clapping extensions must remain separate factors.

\subsection{Study scope}
The evaluation covers one concert, one camera arrangement, one realisation per automatic method and three experts. It therefore characterises this application rather than establishing a general ranking of editing strategies. Further concerts, repeated random realisations and a broader expert or audience study are needed to assess transfer across material and preferences.

\section{Conclusion}
The framework combines audio and visual analysis, explicit camera scoring, beat-based timing and probabilistic selection in a modular editing system. With effectively matched automatic timing, the prediction-driven edit receives a less favourable overall grade than the available random edit from all three raters. The annotation-driven edit improves on predictions for two raters without receiving a better grade than random selection. The result motivates further development of the selection objective and policy.

The component results and expert feedback support a research agenda that extends beyond detector accuracy. The findings motivate testing musical relevance, suitable moments, framing and motivated reactions as distinct factors; quality labels need a defined operational effect, and eligibility, aggregation and sequence context should be independently testable. The component-to-evaluation mapping connects the scoring rules to concrete directions for controlled comparison. A follow-up can therefore investigate how changes propagate from labels to scores, selected views and complete-edit judgement, while retaining both favourable and critical observations.

\section*{Declarations}
\noindent\textbf{Funding.} This work was supported in part by the German Federal Ministry of Education and Research through the DATIpilot programme.
% AUTHORCHECK: Confirm the formal funding wording and add the grant number if required.

\noindent\textbf{Acknowledgements.} The authors thank Anja Hartmann and Boris Terpinc for their expertise in video editing, Jerg Bengel for supervising the reference edit, and Valmire Nivokazi, Sara Kemmler, Nina Floeter and Ali Senpinar for contributing to it. They thank Thorsten Klaes and Heiko Raiser for hardware and livestream support, and acknowledge the preparatory student work of Areej Jumaa, Celine Schmitt, Fadi Altobaji, Lennart Daum\"uller, Min Xu and Nina Kretschmer. The Uli Gutscher Quartet granted permission to use the concert recording.

\noindent\textbf{AI assistance.} AI tools assisted with manuscript editing, organisation, LaTeX and figure preparation, numerical calculations and descriptive grouping of written feedback. The authors are responsible for the scientific content and final submitted version.
% AUTHORCHECK: Confirm funding, author contributions, competing interests and applicable consent/ethics declarations before submission.

\FloatBarrier
\begingroup
\fontsize{9}{10.8}\selectfont
\setlength{\bibsep}{2pt}
\interlinepenalty=10000
\bibliographystyle{plainnat}
\bibliography{references}

@inproceedings{mcfee2015librosa,
	author    = {Brian McFee and Colin Raffel and Dawen Liang and Daniel P. W. Ellis and Matt McVicar and Eric Battenberg and Oriol Nieto},
	title     = {{librosa}: Audio and Music Signal Analysis in Python},
	booktitle = {Proceedings of the 14th Python in Science Conference (SciPy 2015)},
	year      = {2015},
	pages     = {18--24}
}

@inproceedings{objects365,
	author    = {Shuai Shao and Zeming Li and Tianyuan Zhang and Chao Peng and Gang Yu and Xiangyu Zhang and Jing Li and Jian Sun},
	title     = {{Objects365}: A Large-Scale, High-Quality Dataset for Object Detection},
	booktitle = {Proceedings of the IEEE/CVF International Conference on Computer Vision (ICCV)},
	year      = {2019},
	pages     = {8429--8438},
	doi       = {10.1109/ICCV.2019.00852}
}

@InProceedings{Wang2023,
  author     = {Wang, Caihong and Qiu, Chuhan and Xu, Wanxin and Zhu, Weijie and Luo, Hanlu and Chen, Lei and Yu, Ziyi and Fu, Lin and Chen, Xuanjun},
  booktitle  = {Proceedings of the 2023 ACM International Conference on Interactive Media Experiences Workshops},
  title      = {Intelligent Directing System for Music Concert Scene Based on Visual and Auditory Information},
  year       = {2023},
  month      = jun,
  pages      = {95--102},
  publisher  = {ACM},
  series     = {IMXw '23},
  collection = {IMXw '23},
  doi        = {10.1145/3604321.3604336},
}

@inproceedings{Peng2024,
  title = {{D-FINE}: Redefine Regression Task in {DETRs} as Fine-grained Distribution Refinement},
  author = {Yansong Peng and Hebei Li and Peixi Wu and Yueyi Zhang and Xiaoyan Sun and Feng Wu},
  booktitle = {International Conference on Learning Representations},
  year = {2025},
  url = {https://proceedings.iclr.cc/paper_files/paper/2025/file/6cf58a87e3097e7d1f9be3e8693a93de-Paper-Conference.pdf}
}

@book{gonzalez2018,
  author    = {Rafael C. Gonzalez and Richard E. Woods},
  title     = {Digital Image Processing},
  edition   = {4},
  publisher = {Pearson},
  year      = {2018}
}

@INPROCEEDINGS{Daniyal2011,
  author={Daniyal, Fahad and Cavallaro, Andrea},
  booktitle={2011 Conference for Visual Media Production}, 
  title={Multi-camera Scheduling for Video Production}, 
  year={2011},
  volume={},
  number={},
  pages={11-20},
  doi={10.1109/CVMP.2011.8}}

@article{Rockafellar2000,
  author  = {R. Tyrrell Rockafellar and Stanislav Uryasev},
  title   = {Optimization of Conditional Value-at-Risk},
  journal = {Journal of Risk},
  volume  = {2},
  number  = {3},
  pages   = {21--41},
  year    = {2000}
}

@ARTICLE{Stoll2023,
  author={Stoll, Eckhard and Breide, Stephan and G{\"o}ring, Steve and Raake, Alexander},
  journal={IEEE Access}, 
  title={Automatic Camera Selection, Shot Size, and Video Editing in Theater Multi-Camera Recordings}, 
  year={2023},
  volume={11},
  number={},
  pages={96673-96692},
  doi={10.1109/ACCESS.2023.3311256}}

@inproceedings{Shrestha2010,
author = {Shrestha, Prarthana and de With, Peter H.N. and Weda, Hans and Barbieri, Mauro and Aarts, Emile H.L.},
title = {Automatic mashup generation from multiple-camera concert recordings},
year = {2010},
isbn = {9781605589336},
publisher = {Association for Computing Machinery},
address = {New York, NY, USA},
url = {https://doi.org/10.1145/1873951.1874023},
doi = {10.1145/1873951.1874023},
pages = {541--550},
numpages = {10},
location = {Firenze, Italy},
series = {MM '10}
,
booktitle = {Proceedings of the 18th ACM International Conference on Multimedia}
}

@inproceedings{Liu2001,
author = {Liu, Qiong and Rui, Yong and Gupta, Anoop and Cadiz, J. J.},
title = {Automating camera management for lecture room environments},
year = {2001},
isbn = {1581133278},
publisher = {Association for Computing Machinery},
address = {New York, NY, USA},
url = {https://doi.org/10.1145/365024.365310},
doi = {10.1145/365024.365310},
booktitle = {Proceedings of the SIGCHI Conference on Human Factors in Computing Systems},
pages = {442--449},
numpages = {8},
location = {Seattle, Washington, USA},
series = {CHI '01}
}

@article{Arev2014,
author = {Arev, Ido and Park, Hyun Soo and Sheikh, Yaser and Hodgins, Jessica and Shamir, Ariel},
title = {Automatic editing of footage from multiple social cameras},
year = {2014},
issue_date = {July 2014},
publisher = {Association for Computing Machinery},
address = {New York, NY, USA},
volume = {33},
number = {4},
issn = {0730-0301},
url = {https://doi.org/10.1145/2601097.2601198},
doi = {10.1145/2601097.2601198},
journal = {ACM Trans. Graph.},
month = jul,
articleno = {81},
numpages = {11}
}

@inproceedings{GonzalbezBiosca2025,
  author = {Daniel Gonz{\'a}lbez-Biosca and Josep Cabacas-Maso and Carles Ventura and Ismael Benito-Altamirano},
  title = {When and How to Cut Classical Concerts? A Multimodal Automated Video Editing Approach},
  booktitle = {Proceedings of the 3rd International Workshop on Multimedia Content Generation and Evaluation: New Methods and Practice},
  series = {McGE '25},
  year = {2025},
  pages = {98--106},
  publisher = {Association for Computing Machinery},
  doi = {10.1145/3746278.3759387},
  url = {https://doi.org/10.1145/3746278.3759387}
}

@inproceedings{Jimenez2022,
  author = {Albert Jim{\'e}nez and Llu{\'i}s G{\'o}mez and Joan Llobera},
  title = {Automated Video Edition for Synchronized Mobile Recordings of Concerts},
  booktitle = {Proceedings of the 17th International Joint Conference on Computer Vision, Imaging and Computer Graphics Theory and Applications},
  volume = {4: VISAPP},
  year = {2022},
  pages = {941--948},
  publisher = {SCITEPRESS},
  doi = {10.5220/0010847600003124},
  url = {https://doi.org/10.5220/0010847600003124}
}

@misc{Rao2022,
  author = {Anyi Rao and Xuekun Jiang and Sichen Wang and Yuwei Guo and Zihao Liu and Bo Dai and Long Pang and Xiaoyu Wu and Dahua Lin and Libiao Jin},
  title = {Temporal and Contextual Transformer for Multi-Camera Editing of {TV} Shows},
  year = {2022},
  eprint = {2210.08737},
  archivePrefix = {arXiv},
  primaryClass = {cs.CV},
  doi = {10.48550/arXiv.2210.08737},
  url = {https://arxiv.org/abs/2210.08737},
  note = {arXiv preprint}
}

@inproceedings{CabacasMaso2026,
  author = {Josep Cabacas-Maso and Carles Ventura and Ismael Benito-Altamirano},
  title = {A Dual-Transformer Architecture with Cross-Attention for Multi-Camera View Recommendation},
  booktitle = {Computer Vision -- ECCV 2026},
  series = {Lecture Notes in Computer Science},
  volume = {17059},
  year = {2026},
  pages = {460--475},
  publisher = {Springer},
  address = {Cham},
  doi = {10.1007/978-3-032-37086-0_25},
  url = {https://doi.org/10.1007/978-3-032-37086-0_25}
}

@article{Zhang2024EmotionCast,
  author = {Xinyi Zhang and Xinran Ba and Feng Hu and Jin Yuan},
  title = {{EmotionCast}: An Emotion-Driven Intelligent Broadcasting System for Dynamic Camera Switching},
  journal = {Sensors},
  year = {2024},
  volume = {24},
  number = {16},
  pages = {5401},
  doi = {10.3390/s24165401},
  url = {https://doi.org/10.3390/s24165401}
}

@inproceedings{Galvane2015,
  author = {Galvane, Quentin and Ronfard, R{\'e}mi and Lino, Christophe and Christie, Marc},
  title = {Continuity Editing for {3D} Animation},
  booktitle = {Proceedings of the Twenty-Ninth AAAI Conference on Artificial Intelligence},
  year = {2015},
  pages = {753--761},
  publisher = {AAAI Press},
  doi = {10.1609/aaai.v29i1.9288}
}

@inproceedings{Girmaji2025EditIQ,
  author = {Girmaji, Rohit and Beri, Bhav and Subramanian, Ramanathan and Gandhi, Vineet},
  title = {{EditIQ}: Automated Cinematic Editing of Static Wide-Angle Videos via Dialogue Interpretation and Saliency Cues},
  booktitle = {Proceedings of the 30th International Conference on Intelligent User Interfaces},
  year = {2025},
  publisher = {Association for Computing Machinery},
  doi = {10.1145/3708359.3712113},
  url = {https://arxiv.org/abs/2502.02172}
}
\endgroup
\clearpage
\appendix
\numberwithin{equation}{section}
\numberwithin{table}{section}
\numberwithin{figure}{section}
% Technical input to main.tex; no standalone document or bibliography.
\section{Implementation details}
\label{app:implementation}
The following definitions specify the audio, visual and selection modules. The camera mapping and class supports are given in Table~\ref{tab:cameras}.

\subsection{Beat grid and target shot duration}
\label{app:sec:timing}
The named audio stems are mixed to form the timing waveform, which is loaded as mono at its existing sampling rate using librosa \citep{mcfee2015librosa}. If the WAV file is absent, the cutter extracts a 48\,kHz stereo PCM waveform from CAM4 and then loads it as mono. The beat tracker returns times $\mathcal B=\{b_0,b_1,\ldots\}$. We construct an approximate bar grid
\begin{equation}
\mathcal D=\{b_0,b_4,b_8,\ldots\}.
\end{equation}
This assumes four beats per bar and uses the first detected beat as the phase reference. It does not estimate metre or identify true downbeats. The cutter contains no additional median filter on the beat sequence, and no beat- or bar-annotation accuracy is reported.

Local tempo is estimated over 30\,s windows with a 10\,s hop. At shot start $t$, the implementation uses the first overlapping tempo window in chronological order, with a middle-tempo fallback when no window covers $t$. Its tempo classes and target dwell-time ranges are
\begin{equation}
\Delta\sim\mathcal U(a_f,b_f),\qquad
(a_f,b_f)=
\begin{cases}
(10,15)\ \mathrm{s},& f<90\ \mathrm{BPM},\\
(6,10)\ \mathrm{s},&90\leq f<130\ \mathrm{BPM},\\
(3,6)\ \mathrm{s},&f\geq130\ \mathrm{BPM}.
\end{cases}
\label{app:eq:duration}
\end{equation}
The next boundary is the first grid point at or after $t+\Delta$,
\begin{equation}
t'=\min\{d\in\mathcal D:d\geq t+\Delta\},
\label{app:eq:boundary}
\end{equation}
or the clip end if no such point remains. The clip end is limited by the audio duration and the largest score timestamp. Equation~\ref{app:eq:duration} bounds the sampled target, not the realised shot duration, which can be longer after grid alignment. Beat alignment therefore provides a timing convention, rather than a demonstrated model of phrasing or dramaturgy.

\subsection{Activity and role estimation from audio stems}
\label{app:sec:audio}
Separate stems permit direct energy analysis without identifying instruments from a polyphonic mixture. Stem identities are supplied by the track names. Each stem is loaded at its native sampling rate $f_s$ and analysed with a centred 3\,s RMS frame and a 1.5\,s hop. For stem $i$ and frame $k$,
\begin{equation}
r_i[k]=\sqrt{\frac{1}{N}\sum_{n=0}^{N-1}x_i[kH+n-\lfloor N/2\rfloor]^2},
\quad N=\operatorname{round}(3f_s),\quad H=\operatorname{round}(1.5f_s),
\end{equation}
with zero padding at signal boundaries. A centred nine-frame moving average produces $\bar r_i[k]$. Its effective temporal support is approximately $3+8\times1.5=15$\,s. This smoothing can obscure short entrances and contributes future context, so the role estimator is also an offline procedure.

\begin{samepage}
Normalisation uses shared bounds over all smoothed stem energies, $L=\operatorname{perc}_1(\{\bar r_i[k]\}_{i,k})$ and $U=\operatorname{perc}_{99}(\{\bar r_i[k]\}_{i,k})$, to define the activity:
\begin{equation}
v_i[k]=\clip_{[0,1]}\left(\frac{\bar r_i[k]-L}{U-L}\right).
\label{app:eq:norm}
\end{equation}
\end{samepage}
If the bounds are invalid or $U\leq L$, the implementation uses $L=0$, $U=1$. These are global bounds, not separate percentile scales for individual instruments. Relative stem levels therefore remain dependent on recording gain and microphone bleed.

With $v_{\max}[k]=\max_i v_i[k]$, active stems satisfy both an absolute and a relative threshold:
\begin{equation}
A[k]=\{i:v_i[k]\geq0.075\ \text{and}\ v_i[k]\geq0.35v_{\max}[k]\}.
\end{equation}
For a nonempty active set, normalised energy shares are
\begin{equation}
\alpha_i[k]=\frac{v_i[k]^2}{\sum_{j\in A[k]}v_j[k]^2}.
\end{equation}
A stem is primary if $\alpha_i[k]\geq0.92\max_{j\in A[k]}\alpha_j[k]$; other active stems are secondary. Multiple primary stems are possible. Consecutive equal roles are merged into intervals on the 1.5\,s grid. The export maps announcement, trombone, guitar, bass-pickup and piano stems to their semantic labels, and maps both bass-drum and snare stems to Drums. Roles are assigned before this label mapping, so the exported primary and secondary sets can overlap for a grouped instrument. Labels are cropped to 480\,s. These operational roles describe relative signal energy and are not equivalent to perceptual soloist identification.

\subsection{Visible instruments and image sharpness}
Visual instrument detection uses the pretrained D-FINE-X Objects365 detector \citep{Peng2024,objects365}, with confidence threshold 0.55. A detected instrument is visually primary when its box covers more than 20\% of the frame area. The retained classes are Drums, Guitar, Piano, Cello and Trombone; Cello is mapped to Bass as an application-specific proxy. The detector is used without fine-tuning.

A detected bounding box $(x_1,y_1,x_2,y_2)$ in a frame of width $W$ and height $H_v$ is visually primary if
\begin{equation}
\frac{(x_2-x_1)(y_2-y_1)}{WH_v}>0.20;
\end{equation}
otherwise it is secondary. This is a size proxy; it does not evaluate occlusion, face visibility, shot aesthetics or whether the instrument is being played. The detector is not fine-tuned on the concert, and the visual timeline in Figure~\ref{app:fig:roles} is an example rather than a quantitative detection-accuracy test.

For blur estimation, frames are sampled approximately every 0.2\,s using a frame step of $\max(1,\operatorname{round}(0.2\,\mathrm{fps}))$. Grayscale frames at their input resolution are smoothed with a $5\times5$ Gaussian kernel (OpenCV's automatic sigma setting). Sobel derivatives with kernel size 3 produce $g_x$ and $g_y$ \citep{gonzalez2018}. The sharpness statistic is the population standard deviation over pixels,
\begin{equation}
B_c(t)=\operatorname{std}_{x,y}\!\left(\sqrt{g_x(x,y,t)^2+g_y(x,y,t)^2}\right).
\end{equation}
Frames are fully blurred for $B_c<20$, partially blurred for $20\leq B_c<25$, and otherwise not blurred. Consecutive identical labels are merged into half-open intervals. Because the statistic is not scale-normalised, the thresholds depend on image resolution, texture, exposure and viewpoint. They should not be interpreted as universal focus thresholds.

\subsection{Camera utility from role matches and penalties}
\label{app:sec:utility}
Let $A_p(t)$ and $A_s(t)$ be the global primary and secondary audible label sets, and let $V_{c,p}(t)$ and $V_{c,s}(t)$ be the corresponding visible sets for camera $c$. Scores are evaluated at integer seconds. An interval label is present when $s\leq t<e$. Identical audio intervals repeated across camera tasks are deduplicated. The scoring parser combines Speech and Singing into the token \texttt{Speech/Singing}; clapping labels, when present, are treated as primary event labels.

For audible role $r\in\{p,s\}$ and label $a$, define the match contribution
\begin{equation}
M_{c,r}(a,t)=
\begin{cases}
w_{r,p},&a\in V_{c,p}(t),\\
w_{r,s},&a\in V_{c,s}(t)\setminus V_{c,p}(t),\\
0,&\text{otherwise},
\end{cases}
\end{equation}
where visible primary membership takes precedence if both visible roles occur. The match bonus is $R_c(t)=\sum_{r\in\{p,s\}}\sum_{a\in A_r(t)}M_{c,r}(a,t)$. With $V_c=V_{c,p}\cup V_{c,s}$ and $A=A_p\cup A_s$, the positive penalty magnitude is
\begin{equation}
P_c(t)=0.9\,|A_p(t)\setminus V_c(t)|
+0.5\,|A_s(t)\setminus V_c(t)|
+0.75\,|V_c(t)\setminus A(t)|.
\label{app:eq:penalty}
\end{equation}
The per-second score is
\begin{equation}
S_c(t)=\operatorname{round}_2\!\left([4+R_c(t)-P_c(t)]\,Q_c(t)\right),
\label{app:eq:score}
\end{equation}
with the role weights and quality multipliers in Table~\ref{app:tab:parameters}. The offset does not guarantee nonnegative scores; nonnegativity is enforced at interval aggregation. Partially blurred footage receives no additional score reduction in this configuration. A fully blurred label or a recognised lowercase \texttt{pan} label sets the score to zero. The prediction-driven variant uses blur and instrument-role estimates; the annotation-driven variant also supplies pan and clapping labels.

\begin{table}[htbp]
\centering\small
\caption{System parameters for audio analysis, visual cues, camera scoring and selection.}
\label{app:tab:parameters}
\begin{tabularx}{\textwidth}{@{}Yll@{}}
\toprule
Setting & Symbol / condition & Value\\
\midrule
Visual detector & D-FINE / Objects365 & X\\
Detection confidence & retained predictions & 0.55\\
Primary visual role & fraction of frame area & $>0.20$\\
Audio RMS frame / hop & seconds & 3.0 / 1.5\\
Audio smoothing & centred RMS frames & 9\\
Activity normalisation & shared percentiles & 1 / 99\\
Absolute / relative activity & thresholds & 0.075 / 0.35\\
Primary audio role & relative energy share & $\geq0.92$\\
Sharpness sampling & seconds & 0.2\\
Sharpness filters & Gaussian / Sobel kernels & $5\times5$ / 3\\
Blur classes & full / partial thresholds & $<20$ / $<25$\\
Tempo estimation & window / hop (s) & 30 / 10\\
Tempo classes & BPM boundaries & 90 / 130\\
Target dwell times & slow / middle / fast (s) & 10--15 / 6--10 / 3--6\\
Timing grid & tracked-beat step & 4\\
\midrule
Primary audio, primary visible & $w_{p,p}$ & 4.0\\
Primary audio, secondary visible & $w_{p,s}$ & 1.0\\
Secondary audio, primary visible & $w_{s,p}$ & 1.0\\
Secondary audio, secondary visible & $w_{s,s}$ & 0.5\\
Missing primary audible label & penalty per label & 0.9\\
Missing secondary audible label & penalty per label & 0.5\\
Visible but inaudible label & penalty per label & 0.75\\
Score offset & before quality multiplication & 4.0\\
Quality multipliers & fully / partly blurred / pan & 0 / 1 / 0\\
Lower-tail fraction & $\beta$ & 0.10\\
Transition weights & diagonal / off-diagonal & 0 / 0.30\\
Selection and repeat constraint & categorical / no repeat & enabled\\
Rank conversion & use\_rank\_weights & disabled\\
Score time offset & seconds & 0.0\\
Initial previous-camera state & before first selection & CAM1\\
\bottomrule
\end{tabularx}
\end{table}

\subsection{Lower-tail interval aggregation and camera sampling}
\label{app:sec:selection}
For candidate interval $[t,t')$, the code uses the integer-second bins from $\lfloor t\rfloor$ through $\lceil t'\rceil-1$. This includes the bin containing a fractional start time. Let $n$ be their number and let $S_{c,(1)}\leq\cdots\leq S_{c,(n)}$ denote the sorted scores. The interval utility is the clipped mean of the lowest $m$ samples,
\begin{equation}
q_c([t,t'))=\max\!\left(0,\frac{1}{m}\sum_{j=1}^{m}S_{c,(j)}\right),
\qquad m=\max(1,\lceil\beta n\rceil),\quad\beta=0.10.
\label{app:eq:tail}
\end{equation}
Missing score bins are zero. This empirical lower-tail average is related to lower-tail conditional value-at-risk \citep{Rockafellar2000}; the ceiling convention is the implemented finite-sample rule, rather than an exact fractional-tail estimator. For $n\leq10$, it reduces to the minimum score clipped at zero. This makes brief failures influential, which can reject an otherwise useful shot; the aggregation is a conservative choice, not a proven editorial optimum.

Define $w_c=q_c/\sum_jq_j$ when the sum is positive and $w_c=1/4$ otherwise. Given previous camera $i$, the implementation forms
\begin{equation}
\widetilde p_j=T_{ij}w_j,\qquad
p_j=\frac{\widetilde p_j}{\sum_k\widetilde p_k},\qquad
T_{ij}=\begin{cases}0,&i=j,\\0.30,&i\ne j.\end{cases}
\label{app:eq:selection}
\end{equation}
When all remaining weights vanish, the fallback is uniform over the three cameras other than $i$. The next camera is drawn from this categorical distribution for the entire interval, and becomes the previous camera for the next decision. The initial previous-camera state is CAM1, so CAM1 is excluded from the first draw as well.

The off-diagonal value 0.30 is an unnormalised weight: the final normalisation makes it equivalent to 1/3. With positive eligible utilities, Equation~\ref{app:eq:selection} simplifies to $p_j=q_j/\sum_{k\ne i}q_k$ for $j\ne i$. Thus the transition mechanism is a first-order no-repeat constraint, not a learned model of editing grammar. Scheduling methods can use richer transition models \citep{Daniyal2011}; no such benefit is established here. Moreover, a zero utility is not an unconditional shot veto because the all-zero fallback may still choose an unsuitable view.

Speech and singing share a combined event label. The timing module applies the same policy throughout the clip; a separate announcement policy is a direction for further development.

\FloatBarrier
\section{Component evaluation}
\label{app:component-details}
\label{app:sec:components}
\subsection{Sampling and metric definitions}
Audio labels are evaluated on integer bins $\{\lfloor s\rfloor,\ldots,\lfloor e\rfloor-1\}$ for each interval $[s,e)$ over timestamps 0--480. This yields 481 samples, including a terminal empty bin. Repeated audio labels across camera views form global audible-role sets.

Primary and secondary roles are evaluated as separate multilabel tasks over the six instruments. Not active is the complement of the union of these roles, per instrument. Clapping is outside this six-instrument evaluation. For each label, precision $P$, recall $R$ and F1 are
\begin{equation}
P=\frac{\mathrm{TP}}{\mathrm{TP}+\mathrm{FP}},\qquad
R=\frac{\mathrm{TP}}{\mathrm{TP}+\mathrm{FN}},\qquad
F1=\frac{2\mathrm{TP}}{2\mathrm{TP}+\mathrm{FP}+\mathrm{FN}}.
\end{equation}
Support $N_\ell$ is the number of positive reference samples. Weighted summaries are $\sum_\ell N_\ell m_\ell/\sum_\ell N_\ell$ for metric $m$. They are not micro-averages, and weighted F1 is not calculated from weighted precision and recall. Multilabel supports can exceed the number of time samples. Undefined metrics are set to zero in the weighted calculation; classes absent from the reference are shown with dashes in the table.

For blur, each camera is evaluated at integer instants $s\leq t<e$, with fully blurred labels taking precedence over partially blurred labels. Unlabelled instants are treated as not blurred. The horizon is the maximum interval end in either export, giving the camera-specific sample counts in Table~\ref{tab:cameras}. The metrics describe this recording; adjacent temporal samples are dependent.

\subsection{Audio roles}
\begin{table}[htbp]
\centering\footnotesize
\setlength{\tabcolsep}{2.7pt}
\caption{Audio-role classification against reference labels. Each group reports reference support $N$, precision, recall and F1. The 481 time samples form a multilabel task. Dashes indicate absent reference classes ($N=0$); the underlying evaluator assigns zero to undefined scores.}
\label{app:tab:audio}
\begin{tabular}{@{}lrrrrrrrrrrrr@{}}
\toprule
&\multicolumn{4}{c}{Not active}&\multicolumn{4}{c}{Primary}&\multicolumn{4}{c}{Secondary}\\
\cmidrule(lr){2-5}\cmidrule(lr){6-9}\cmidrule(lr){10-13}
Instrument & $N$ & P & R & F1 & $N$ & P & R & F1 & $N$ & P & R & F1\\
\midrule
Bass &189&.639&1.000&.779&0&\NA&\NA&\NA&292&1.000&.599&.749\\
Drums &189&.508&1.000&.674&95&.667&.021&.041&197&.541&.299&.386\\
Guitar &154&.726&1.000&.842&95&.528&.905&.667&232&.953&.435&.598\\
Piano &481&1.000&.981&.991&0&\NA&\NA&\NA&0&\NA&\NA&\NA\\
Speech &413&.998&.983&.990&68&.905&.985&.944&0&\NA&\NA&\NA\\
Trombone &206&.787&.879&.830&275&.954&.673&.789&0&\NA&\NA&\NA\\
\midrule
Weighted &1632&.848&.975&.895&533&.820&.638&.654&721&.859&.465&.601\\
\bottomrule
\end{tabular}
\end{table}

\begin{figure}[htbp]
\centering
\includegraphics[width=\textwidth]{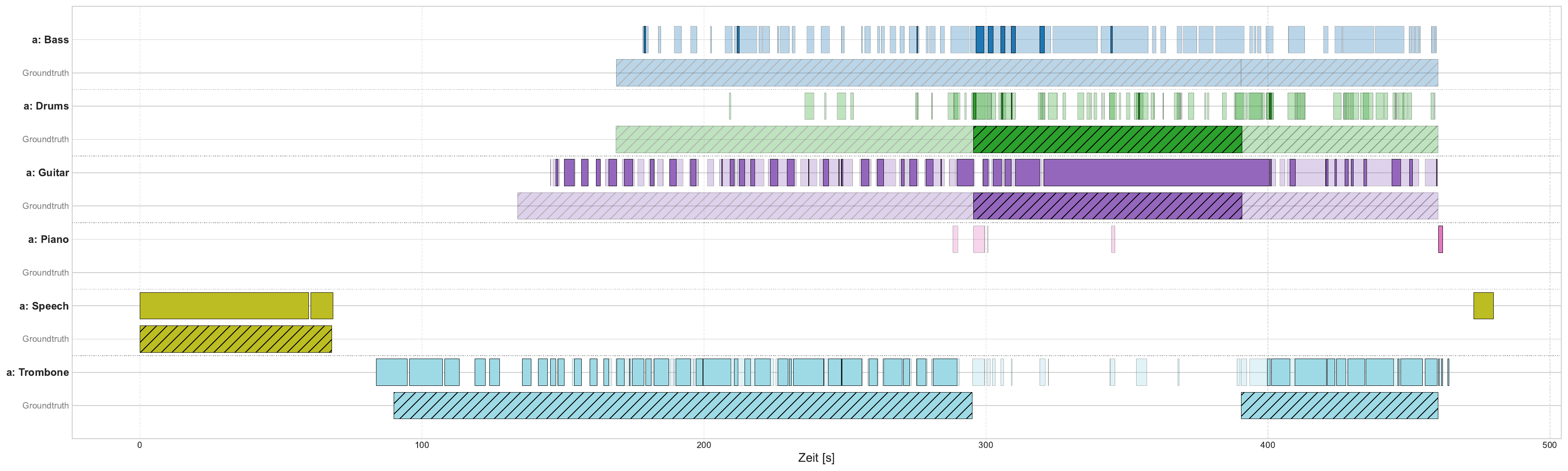}\\[7pt]
\includegraphics[width=\textwidth]{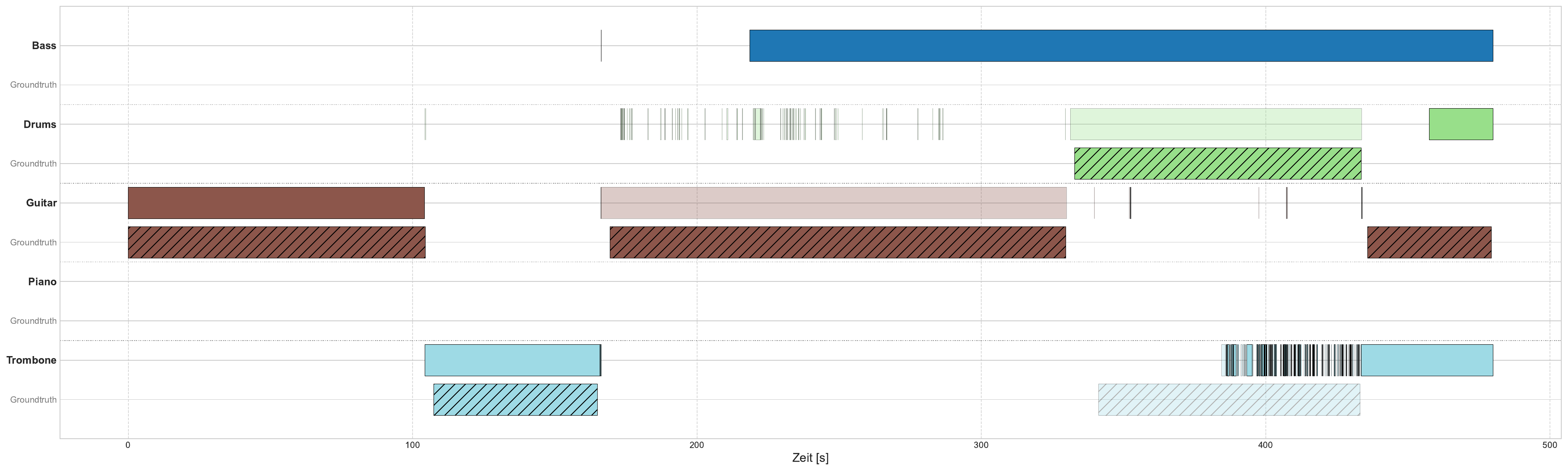}
\caption{Audio-role timelines (top, global labels) and visible instrument roles (bottom, CAM3). Coloured rows are predictions; hatched rows are reference annotations. Saturated colours denote primary roles and pale colours secondary roles. The original axis label ``Zeit [s]'' means time in seconds. The plots illustrate this case and do not establish visual detector accuracy.}
\label{app:fig:roles}
\end{figure}

Table~\ref{app:tab:audio} shows uneven role detection. Speech primary labels have F1 0.944, while primary Drums have recall 0.021 and F1 0.041 despite 95 positive reference samples. Guitar primary recall is 0.905 but precision is 0.528; Trombone primary precision is 0.954 with recall 0.673. The zero reference support for primary Bass and for all active Piano roles prevents claims about detecting those roles. Inactive-role and active-role performance therefore need separate interpretation. Figure~\ref{app:fig:roles} illustrates the fragmentation and timing differences that remain after smoothing.

\subsection{Blur classification}
\begin{table}[htbp]
\centering\footnotesize
\setlength{\tabcolsep}{3.2pt}
\caption{Blur-classification precision, recall and F1. Reference supports and camera-specific $n$ are given in Table~\ref{tab:cameras}. Dashes denote absent reference classes. Weighted summaries use class supports, including the poor results on CAM1.}
\label{app:tab:blur}
\begin{tabular}{@{}lrrrrrrrrrrrr@{}}
\toprule
&\multicolumn{3}{c}{CAM1}&\multicolumn{3}{c}{CAM2}&\multicolumn{3}{c}{CAM3}&\multicolumn{3}{c}{CAM4}\\
\cmidrule(lr){2-4}\cmidrule(lr){5-7}\cmidrule(lr){8-10}\cmidrule(lr){11-13}
Class & P & R & F1 & P & R & F1 & P & R & F1 & P & R & F1\\
\midrule
Not blurred &.01&1.00&.03&.16&.98&.28&.01&1.00&.02&1.00&1.00&1.00\\
Partially blurred &.00&.00&.00&.90&.25&.39&.94&.30&.46&\NA&\NA&\NA\\
Fully blurred &\NA&\NA&\NA&.38&.45&.41&.44&.29&.35&\NA&\NA&\NA\\
\midrule
Weighted &.00&.01&.00&.77&.35&.38&.91&.31&.45&1.00&1.00&1.00\\
\bottomrule
\end{tabular}
\end{table}

The partially blurred class on CAM2 and CAM3 has high precision but low recall; fully blurred detections are weaker. CAM1 is predicted as not blurred throughout although 474 of its 481 reference samples are partially blurred, producing near-zero weighted F1. CAM4 contains only not-blurred reference samples, so performance on blurred frames is not assessed for that view. Fixed, unnormalised image-gradient thresholds are particularly vulnerable to camera-dependent appearance. Moreover, the scoring configuration does not penalise partial blur, limiting the impact of this classification even when it is correct.

\begin{figure}[htbp]
\centering
\includegraphics[width=\textwidth]{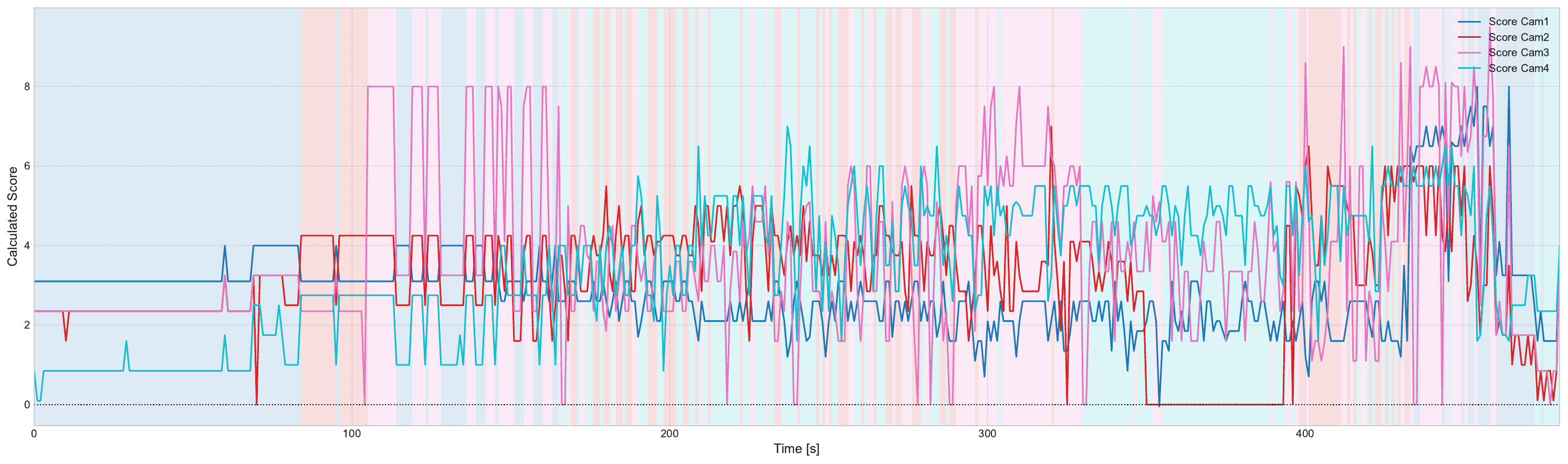}\\[7pt]
\includegraphics[width=\textwidth]{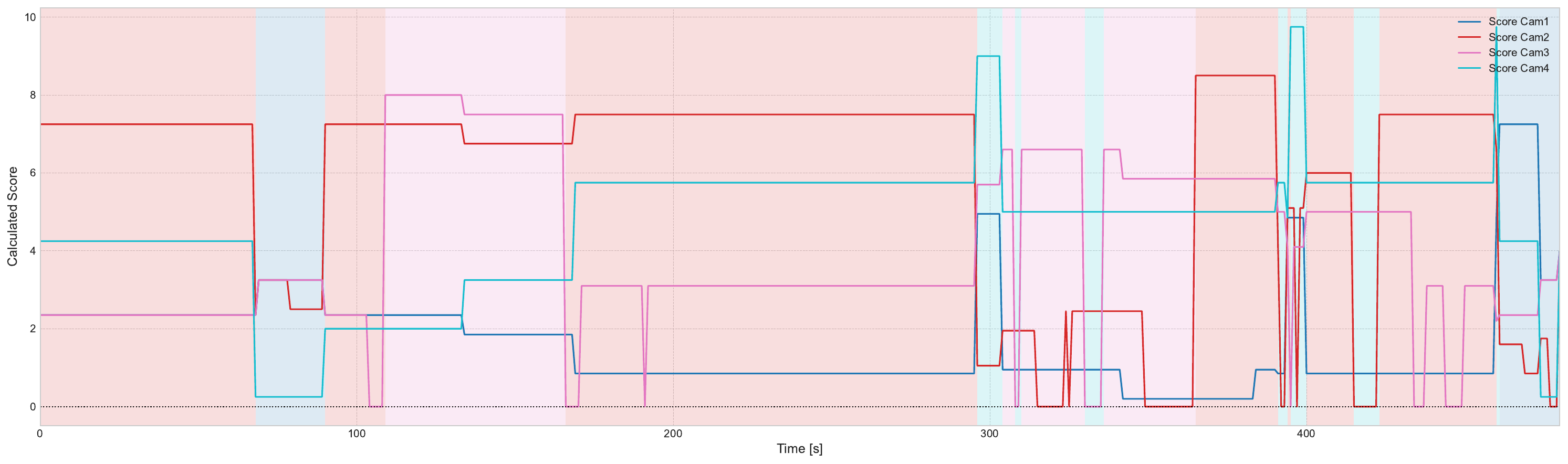}
\caption{Per-second camera scores from predicted labels (top) and reference annotations (bottom). Background shading indicates the instantaneous highest-scoring camera, while the selector samples cameras over complete intervals. The annotation-driven scores additionally use pan/clapping cues.}
\label{app:fig:scores}
\end{figure}

Figure~\ref{app:fig:scores} compares the resulting camera-score profiles from predicted and reference labels.

\FloatBarrier
\section{Additional temporal documentation}

\label{app:cuts}
For a window $W=[a,b)$ and successive video boundaries $c_j,c_{j+1}$, the positive displayed segments are
\begin{equation}
\ell_j(W)=\max\{0,\min(c_{j+1},b)-\max(c_j,a)\}.
\end{equation}
The sample SD is $s=\sqrt{\sum_j(\ell_j-\bar\ell)^2/(n-1)}$, undefined for $n=1$. Boundary clipping is distinct from a new camera cut.

Table~\ref{app:tab:phaseappendix} supplements the music-specific analysis with the announcement and applause. Because these windows are approximate and shots crossing their boundaries are clipped, the applause statistics in particular describe short window fragments rather than independent complete shots. Sample SD is undefined for $n=1$, rather than set to zero. The boundary sensitivity rows use only complete shots contained within the music window: 33 for V1 and 54 for each automatic variant. The contrast between longer, more variable manual shots and shorter automatic shots remains descriptively visible; this is not a statistical test of musical appropriateness.

\begin{table}[htbp]
\centering\small
\caption{Additional shot-duration summaries (seconds). Announcement and applause rows include clipped boundary segments; ``Music: complete'' excludes both clipped music-boundary segments. SD is the sample standard deviation.}
\label{app:tab:phaseappendix}
\begin{tabular}{@{}llrrrr@{}}
\toprule
Variant & Window / inclusion & $n$ & Mean & Median & SD\\
\midrule
V1 & Announcement & 1 & 80.00 & 80.00 & \NA\\
V2 & Announcement & 10 & 8.00 & 8.86 & 3.20\\
V3 & Announcement & 10 & 8.00 & 8.86 & 3.21\\
V4 & Announcement & 10 & 8.00 & 8.86 & 3.21\\
\midrule
V1 & Applause & 2 & 3.85 & 3.85 & 4.09\\
V2 & Applause & 2 & 3.85 & 3.85 & 3.99\\
V3 & Applause & 2 & 3.85 & 3.85 & 3.95\\
V4 & Applause & 2 & 3.85 & 3.85 & 3.95\\
\midrule
V1 & Music: complete & 33 & 10.02 & 6.31 & 8.82\\
V2 & Music: complete & 54 & 6.72 & 6.66 & 2.18\\
V3 & Music: complete & 54 & 6.72 & 6.66 & 2.18\\
V4 & Music: complete & 54 & 6.72 & 6.66 & 2.18\\
\bottomrule
\end{tabular}
\end{table}

The four edits contain 230 internal camera changes in total. Cut times are defined by the first frame of the new camera view.

\section{Questionnaire dimensions}
\label{app:questionnaire}
Each video is rated on five dimensions, with four item-specific comment fields and general remarks. Table~\ref{app:tab:questionnaire} summarises the dimensions.

\begin{table}[htbp]
\centering\small
\caption{Dimensions assessed in the expert evaluation.}
\label{app:tab:questionnaire}
\begin{tabularx}{\textwidth}{@{}lY@{}}
\toprule
Field & Assessed dimension\\
\midrule
Ovr & Overall quality and readiness for publication or broadcast.\\
Ed & Appropriateness of cut timing to the musical and content progression.\\
Tr & Coherence and smoothness of image transitions; disruptive changes.\\
Sh & Overall picture quality and sharpness of the selected views.\\
Im & Coherence and aesthetic quality of the overall impression.\\
\bottomrule
\end{tabularx}
\end{table}

\FloatBarrier
\section{Individual expert responses}
\label{app:rawratings}
E1--E3 denote the three experts. The table contains all numeric responses; dashes indicate missing answers.

\begin{table}[htbp]
\centering\small
\caption{All recorded numbers, E1 / E2 / E3. A dash means no numeric answer.}
\label{app:tab:rawratings}
\begin{tabular}{@{}lccccc@{}}
\toprule Variant & Ovr & Ed & Tr & Sh & Im\\\midrule
V1 &3 / 3 / 3&3 / 2 / 2&3 / 5 / 2&2 / 1 / 1&2 / 3 / 3\\
V2 &4 / 3 / 1&4 / 4 / 2&3 / 3 / 1&2 / 1 / 3&3 / 3 / 2\\
V3 &5 / 4 / 3&5 / 4 / \NA&\NA\ / 5 / \NA&\NA\ / 1 / \NA&\NA\ / 4 / \NA\\
V4 &4 / 4 / 1&\NA\ / 4 / 1&\NA\ / 3 / 1&2 / 1 / 2&\NA\ / 4 / 1\\
\bottomrule\end{tabular}
\end{table}
\begin{table}[htbp]\centering\small
\caption{Number of explicit numeric answers. Missing numeric answers remain missing even where a comment refers to another video.}
\begin{tabular}{@{}lrrrrr@{}}\toprule
Variant & Ovr & Ed & Tr & Sh & Im\\\midrule
V1 &3&3&3&3&3\\ V2 &3&3&3&3&3\\
V3 &3&2&1&1&1\\ V4 &3&2&2&3&2\\\bottomrule
\end{tabular}\end{table}
\FloatBarrier

\section{Descriptive analysis of written feedback}
\label{app:qualitative}
Whole comment fields form the units of the descriptive thematic reading. A field can cover more than one theme. Empty fields and references to other videos are distinguished from substantive statements; a cross-reference does not generate another independent observation. The grouping retains positive and negative statements, explicit preferences and disagreements between experts. Themes describe expressed judgements rather than the prevalence of a problem or its technical cause. AI assistance in grouping the comments is disclosed in the main text.

\FloatBarrier
\section{Score sensitivity and camera-selection effects}
\label{app:architecture-tests}
An example illustrates how a change in a visibility label can affect camera scoring and selection. All other labels and the quality state remain fixed. For an audible label that is primary only, changing its visibility from absent in both visible-role sets to visually primary adds the primary--primary bonus of 4 and removes the missing-primary penalty of 0.9. The per-second score therefore increases by 4.9 when $Q_c=1$; when $Q_c=0$, both scores are zero.

Lower-tail aggregation and clipping at zero can leave the interval utility unchanged. The no-repeat rule also excludes the previous camera regardless of its score. For an eligible camera, normalisation depends on the utilities of the other eligible views: if it is the only view with positive utility, its probability remains one despite an increase in utility. Even when probabilities change, categorical sampling can return the same camera. These conditional effects distinguish label changes, score changes and camera choices; none alone establishes improved editorial quality.

\end{document}